\documentclass[%
twocolumn,
preprintnumbers,
superscriptaddress,
nofootinbib,
reprint,
showkeys,
showpacs,
amsmath,amssymb,
aps,
]{revtex4-1}

\usepackage{graphicx}
\usepackage{dcolumn}
\usepackage{bm}
\usepackage{amsfonts,amsbsy}
\usepackage[nodots]{numcompress}
\usepackage{aas_macros}
\usepackage{mathrsfs}
\usepackage{url}
\usepackage{multirow}
\usepackage{xcolor}
\usepackage{color}
\usepackage{ulem}
\usepackage{enumerate}
\usepackage{textcmds}
\usepackage{mathtools}
\usepackage{booktabs}
\usepackage{tabularx}
\usepackage[linkcolor=blue,citecolor=blue,backref=page]{hyperref} 

\renewcommand{\vec}[1]{\boldsymbol{#1}}

\newcommand{\dif}{\mathrm{d}}

\newcolumntype{p}{D{,}{\pm}{-1}}

\begin{document}

\preprint{Version \today}

\title{Excesses of Cosmic Ray Spectra from A Single Nearby Source}

\author{Wei Liu}
 \email{liuwei@ihep.ac.cn}
\author{Xiao-Jun Bi}%
 \email{bixj@ihep.ac.cn}
\author{Su-Jie Lin}
\author{Bing-Bing Wang}
\author{Peng-Fei Yin}
\affiliation{%
Key Laboratory of Particle Astrophysics,
Institute of High Energy Physics, Chinese Academy of Sciences,
Beijing 100049, China
}%

\date{\today}


\begin{abstract}
Growing evidence reveals universal hardening on various cosmic ray spectra, e.g. proton, positron, as well as antiproton fraction. Such universality may indicate they have a common origin. In this paper, we argue that these widespread excesses can be accounted for by a nearby supernova remnant surrounded by a giant molecular cloud. Secondary cosmic rays ($\rm p$, $\rm e^+$) are produced through the collisions between the primary cosmic ray nuclei from this supernova remnant and the molecular gas. Different from the background, which is produced by the ensemble of large amount of sources in the Milky Way, the local injected spectrum can be harder. The time-dependent transport of particles would make the propagated spectrum even harder. Under this scenario, the anomalies of both primary ($\rm p$, $\rm e^-$) and secondary ($\rm e^+$, $\rm \bar{p}/p$) cosmic rays can be properly interpreted. We further show that the TeV to sub-PeV anisotropy of proton is consistent with the observations if the local source is relatively young and lying at the anti-Galactic center direction.

\end{abstract}

\pacs{Valid PACS appear here}


\maketitle

\section{Introduction}
\label{sec:intro}

It is widely accepted that cosmic rays (CRs) below the knee region originate from the galactic supernova remnant (SNR). In light of both primordial diffusive shock acceleration\citep{1983RPPh...46..973D, 1987PhR...154....1B} and steady-state transport\citep{2002astro.ph.12111M, 2007ARNPS..57..285S} processes, the galactic CRs are expected to fall off as a featureless power-law in a wide energy range from tens of GeV to $\sim$ PeV.

However, such a simple picture has been challenged by new observations in the past few years. The anomalies of CR electrons/positrons have been observed by quite a number of experiments, such as HEAT\citep{2001ApJ...559..296D, 2004PhRvL..93x1102B}, AMS-01\citep{2007PhLB..646..145A}, ATIC\citep{2008Natur.456..362C}, PPB-BETS\citep{2008arXiv0809.0760T}, PAMELA\citep{2009Natur.458..607A}, Fermi-LAT\citep{2009PhRvL.102r1101A}, and most recently by AMS-02\citep{2014PhRvL.113l1102A, 2014PhRvL.113l1101A}. The overabundance of positrons requires extra primary sources, which involve either astrophysical objects, such as pulsars\citep{1970ApJ...162L.181S, 2001A&A...368.1063Z, 2009PhRvL.103e1101Y, 2009JCAP...01..025H, 2012CEJPh..10....1P, 2015APh....60....1Y} and the hadronic interactions in SNRs\citep{2009PhRvL.103e1104B, 2009PhRvD..80f3003F, 2009ApJ...700L.170H}, or more exotic origins like the dark matter self-annihilation or decay\citep{2008PhRvD..78j3520B, Cirelli:2008pk, 2009PhLB..672..141B, 2009PhRvD..79b3512Y, 2009PhRvL.103c1103B, 2009PhRvD..80b3007Z}. For an extensive introduction of relevant models, one can refer to the reviews\citep{2009MPLA...24.2139H, 2010IJMPD..19.2011F, 2012APh....39....2S, 2012Prama..79.1021C, 2013FrPhy...8..794B} and references therein.


Observations of proton, helium, and heavier nuclei also show remarkable hardenings at energies above a few hundred GeV/nucleon\citep{2015PhRvL.114q1103A, 2015PhRvL.115u1101A, 2009BRASP..73..564P, 2011Sci...332...69A, 2010ApJ...714L..89A, 2011ApJ...728..122Y}. The hadronic hardening may be ascribed to source model\citep{2010ApJ...725..184B, 2011PhRvD..84d3002Y, 2012A&A...544A..92B, 2012MNRAS.421.1209T, 2013MNRAS.435.2532T, 2013A&A...555A..48B, 2015RAA....15...15L}, transport effect\citep{2012PhRvL.109f1101B, 2012ApJ...752L..13T, 2012JCAP...01..010B, 2012ApJ...752...68V, 2014A&A...567A..33T, 2015arXiv150908227G},  or production mechanism\citep{2011ApJ...729L..13O, 2012PhPl...19h2901M}. Most recently, AMS-02 collaboration released the measurements of antiproton-to-proton ratio, which shows a flat behavior up to $\sim 400$ GeV\citep{2016PhRvL.117i1103A}. This is not expected by the conventional production and propagation pattern of antiprotons\citep{2008JCAP...10..018E, 2015ApJ...803...54K}. Although sizable errorbars at high energies makes this conclusion not overwhelmingly yet, it is worth to chew over the physical implication of a flat $\rm \bar{p}/p$ fraction\citep{2015JCAP...09..023G, 2015arXiv150407230L, 2015PhLB..747..523H, 2015PhRvD..92e5027J, 2015PhRvD..91k1701I, 2015arXiv151004032L}.

In this paper, we aim at relating all these anomalies together in a unified framework. We propose that a local, fresh SNR, which was surrounded by a giant molecular cloud(MC), can explain all the above mentioned phenomena. In this scenario, both primary nuclei and electrons were accelerated by the shock wave in the local SNR, whereas the redundant positrons and antiprotons were yielded through the hadronic interactions between the primary nuclei and molecular gas. Compared to the background component from the ensemble of large amount of Galactic SNRs, the injection spectra of this local SNR is harder. Similar picture has been suggested by \citep{2009PhRvD..80f3003F, 2016PTEP.2016b1E01K}, in which they focused only on the secondaries, namely positrons and antiprotons. Here we further extend the model to study its influence on primary CRs. Moreover, the consequence on Boron-to-Carbon (B/C) ratio is evaluated in a realistic way, with reasonable assumption of the source nuclear abundance. This is especially important because the preliminary data published by AMS-02 collaboration do not show significant deviation from the prediction of standard propagation model\citep{2008JCAP...10..018E}. This, in turn, can impose strong constraints on the models which generate abundant secondary particles\citep{2014PhRvD..89d3013C}. Thirdly, we deduce both background and local SNR parameters coherently without assumptions of the background parameters a priori, which renders our treatment more self-consistent.


The rest of paper is organized as follows: in section \ref{sec:model}, we briefly review the production and propagation of CRs in our model. The results and some discussions are presented in section \ref{sec:res}. Finally in section \ref{sec:concl}, we give our conclusion.

\section{Cosmic Ray Transport Model}
\label{sec:model}

After being injected into the interstellar space, CRs diffuse within the Galactic magnetic halo by scattering off magnetic waves and MHD turbulence. The magnetic halo is usually approximated as a cylinder, with its radial boundary equal to the Galactic radius $R = 20$ kpc. Its half thickness $L$, which characterizes the vertical stretch of interstellar magnetic field, is fixed by CR data. Both CR sources and interstellar medium (ISM) are chiefly distributed in the Galactic disk, whose average width is roughly $200$ pc, much less than halo's thickness. The transport process of CRs in the magnetic halo is described by the following diffusion equation\citep{2007ARNPS..57..285S, 2002astro.ph.12111M}, i.e.
\begin{eqnarray}
\frac{\partial \psi}{\partial t} &=& Q(\vec{r}, p) + \nabla \cdot ( D_{xx}\nabla\psi - \vec{V}_{c}\psi )
+ \frac{\partial}{\partial p}\left[p^2D_{pp}\frac{\partial}{\partial p}\frac{\psi}{p^2}\right]
\nonumber\\
&& - \frac{\partial}{\partial p}\left[ \dot{p}\psi - \frac{p}{3}(\nabla\cdot\vec{V}_c)\psi \right]
- \frac{\psi}{\tau_f} - \frac{\psi}{\tau_r} ~.
\label{propagation_equation}
\end{eqnarray}
Here $\psi(\vec{r}, p, t) = \dif n/\dif p$ is the CR density per total particle momentum $p$ at position $\vec{r}$. At the halo boundary, free escape condition is applied by default, namely $\psi(R, z, p) = \psi(r, \pm L, p) = 0$. The diffusion coefficient $D_{xx}$ is supposed to be isotropic on large scales, which grows with particle rigidity ${\cal R} = pc/Ze$,  viz., 
\begin{equation}
D_{xx} = D_0 \beta  \left( \dfrac{\cal R}{{\cal R}_0} \right)^{\delta} ~,
\end{equation}
where $\beta$ is the particle velocity in unit of the speed of light $c$. Both $D_0$ and $\delta$ are specified as free parameters. Apart from diffusion, CR particles may still suffer from galactic convection $\vec{V}_c$, diffusive reacceleration $D_{pp}$, fragmentation $\tau_f$, radioactive decay $\tau_r$, energy loss $\dot{p}$, etc. In this work, we adopt the diffusive-reacceleration(DR) scenario\citep{2015PhRvD..91f3508L}. The diffusive reacceleration originates from the random motion of magnetic field in ISM, which brings about the second-order Fermi acceleration during the transport. It is usually depicted as the diffusion in momentum space, whose diffusion coefficient $D_{pp}$ is related to spatial diffusion coefficient $D_{xx}$ and Alfv{\'e}nic velocity $v_A$ by
\begin{equation}
D_{pp} D_{xx} = \frac{4p^2v_A^2}{3\delta(4-\delta^2)(4-\delta)} ~.
\end{equation}
 
In this work, the Galactic SNRs are separated into two classes, one is the local fresh SNR and the others are background ones. For the background SNRs, it is reasonable to assume that the spatial distribution of CRs from them arrives at steady state. Nevertheless for the local single SNR, the time-dependent transport of CRs after injection is requisite. In the next subsections, we discuss these two components separately.

\subsection{Background Supernova Remnants} \label{subsec:bkg}

All of the SNRs other than the nearby one are labeled as background sources. Here for simplicity, the spatial distribution of background SNRs is assumed to be axisymmetric, that is, 
\begin{equation}
f(r, z) = \left(\frac{r}{r_{\odot}} \right)^{a} \exp\left[-b\cdot \frac{(r-r_{\odot})}{r_{\odot}} \right] \exp\left[-\frac{|z|}{z_s} \right] ~,
\end{equation}
with $r_{\odot} = 8.5$ kpc, the distance from the solar system to the Galactic center. The parameters $\alpha$ and $\beta$ adjusted to be compatible with the Fermi-LAT gamma-ray data\citep{2011ApJ...729..106T}. The injection spectrum of both primary CR nuclei and electron is paramterized as a broken power-law, namely 
\begin{equation}
  q^{\rm i} =  q^{\rm i}_0 \times\left\{ \begin{array}{ll}
    \left( \dfrac{R}{R_{\rm br}} \right)^{-\nu_1} & R \le R_{\rm br}\\
    \left( \dfrac{R}{R_{\rm br}} \right)^{-\nu_2} & R > R_{\rm br}
  \end{array}
  \right. ~.
\label{inject_spec_nuclei}
\end{equation}

Some other species, e.g. Li, Be, B, ${\rm e}^+$, $\bar{\rm p}$ and radioactive elements, are hardly synthesized during the stellar nucleosynthesis. They can be brought forth from the fragmentation of parent nuclei throughout the transport. For the production of Li, Be and B, usually the so-called straight-ahead approximation is widely used, in which the kinetic energy per nucleon is conserved during the spallation process. The production rate is thus
\begin{equation}
Q_j = \sum_{i = \rm C, N, O} (n_{\rm H} \sigma_{i+{\rm H}\rightarrow j} +n_{\rm He} \sigma_{i+{\rm He} \rightarrow j} ) v \psi_i ~,
\end{equation}
where $n_{{\rm H}/{\rm He}}$ is the number density of hydrogen/helium in the ISM and $\sigma_{i+{\rm H/He}\rightarrow j}$ is the total cross section of the corresponding hadronic interaction. 

Unlike above secondary CR nuclei, secondary $\rm e^+$ and $\rm \bar{p}$ have energy distribution. Therefore the source term of both $\rm e^+$ and $\rm \bar{p}$ is the convolution of the energy spectra of primary nuclei $\Phi_i(E)$ and the relevant differential cross section $d \sigma_{i + {\rm H/He} \to j}/d E_j$, i.e.
 \begin{eqnarray}
\nonumber Q_j &=& \sum_{i = \rm p, He} \int dp_i v \left\lbrace n_{\rm H} \frac{\dif \sigma_{i+{\rm H}\rightarrow j}(p_i, p_j)}{\dif p_j} \right. \\
& & \left. +n_{\rm He} \frac{\dif \sigma_{i+{\rm He}\rightarrow j}(p_i, p_j)}{\dif p_j} \right\rbrace  \psi_i(p_i) ~,
\end{eqnarray}
Furthermore, antiprotons may still undergo non-annihilated inelastic scattering with ISM protons during propagation, in which antiprotons lose a significant amount of their kinetic energies. This is also known as the tertiary production. The numerical package GALPROP\footnote{http://galprop.stanford.edu/} is introduced to solve the transport equation (\ref{propagation_equation}) to obtain the background CRs. The transport parameters are fine-tuned to fit the data together with the contribution from local source.

\subsection{Nearby Young Supernova Remnant}
\label{subsec:nySNR}

We assume a local($< $ kpc) supernova explosion occurred in a giant MC about $10^5-10^6$ years ago. The charged particles were continually accelerated near the shock front with the expansion of supernova ejecta. The accelerated spectrum is represented by a power-law plus an exponential cut-off, i.e. $Q_j = q_j ({\cal R}/{\cal R}_{0})^{-\alpha} \exp[-E/E_{\rm cut}]$. The normalization $q_j$ for each kind of element relies on the local chemical environment. In this work, we suppose that the element abundances in the MC are identical with the average of Galaxy. 

Besides, the CR nuclei generated by the local SNR also collide with the molecular gas around them and give birth to prolific daughter particles, like B, $\rm e^{\pm}$, $\rm \bar{p}$, and so forth. The yields of B and $\rm e^{\pm}$, $\rm \bar{p}$ inside the MC are respectively
\begin{eqnarray}
 Q_j = \sum_{i = \rm C, N, O} (n_{\rm H} \sigma_{i+{\rm H}\rightarrow j} +n_{\rm He} \sigma_{i+{\rm He}\rightarrow j} )v Q_i(E) t_{\rm col}
 \end{eqnarray}
and 
\begin{equation}
\begin{split}
Q_j  \; = \;
\sum_{i = \rm p, He} {\displaystyle \int\limits_{E_{\rm th}}^{+ \infty}} \; d E_i \;
v \; \left\lbrace n_{\, \rm H}
{\displaystyle \frac{d \sigma_{i + {\rm H} \to j}}{d E_j }} \right. \\
\left. +n_{\, \rm He} {\displaystyle \frac{d \sigma_{i + {\rm He} \to j}}{d E_j }} \right\rbrace
Q_i(E_i) t_{\rm col} ~,
\label{sec_source}
\end{split}
\end{equation}
where $n_{\rm H/He}$ is the number density of hydrogen/helium in MC. In this work, we assume that it is $1000$ times greater than the mean value of ISM. $t_{\rm col}$ is the duration of collision. $Q_i(E)$ is the accelerated spectrum of primary nuclei inside local SNR. 

When the radius of shock front is comparable to or even larger than the size of MC, the whole MC was finally fragmented subject to the expansion of ejecta. All the CRs break out of the MC and diffuse into the interstellar space. The time-dependent distribution of CR nuclei from a point source can be obtained by means of Green function technique. Since the local SNR only makes significant contribution above tens of GeV, a simplified transport equation can competent here with other terms, for example reacceleration, energy loss, etc., neglected. Then the transport equation for nuclei is
\begin{equation}
\frac{\partial \varphi_j}{\partial t} -\nabla(D_{xx}\nabla \varphi_j) = Q_j(E)\delta(\vec{r}-\vec{r}^{\prime})\delta(t -t^{\prime}) -2h\delta(z)\Gamma_j \varphi_j ~,
\label{eq:t_dep_diffu_p}
\end{equation}
Here the transport equation is rewritten in terms of $\varphi = \dif n/\dif E$ with $E$ the energy per nucleon. $2h\delta(z) \Gamma_j$ is the fragmentation term and $\Gamma^j_{\rm sp} = (n_{\rm H} \sigma_{j+{\rm H}} +n_{\rm He} \sigma_{j+{\rm He}})v$ denotes the spallation rate of CR nuclei $j$. Meanwhile owing to that the interstellar medium is chiefly concentrated in the Galactic disk, which is much less than halo boundary, thin-disk approximation is applied. The corresponding analytical solution can be found in \citep{2012A&A...544A..92B}. As for the energetic electrons and positrons, the energy loss term, due to synchrotron in interstellar magnetic field and inverse Compton scattering off CMB photons, is far more important here compared with fragmentation. Thus the transport equation for electron and positron becomes
\begin{equation}
\frac{\partial \varphi_j}{\partial t} -D_{xx}\Delta \varphi_j +\frac{\partial}{\partial E}(\dot{E} \varphi_j) = Q_j(E)\delta(\vec{r}-\vec{r}^{\prime})\delta(t -t^{\prime}) ~.
\label{eq:t_dep_diffu_e}
\end{equation}
The energy loss rate $\dot{E}$ is approximately written as $\dot{E} = -bE^2$ in Thomson limit. Here we use the analytical solution available in \citep{2004ApJ...601..340K}.

\section{Results}
\label{sec:res}
In this paper, we apply our model to simultaneously fit the spectra of $\rm \bar{p}/p$, $\rm p$ and $\rm e^{+}$, as well as $\rm B/C$ and $\rm e^{-}$. The local SNR is supposed to lie at $0.1$ kpc away from our solar system. In our propagation model, the essential transport parameters are $D_0$, $\delta$, $v_A$ and $L$. The source parameters of the background CRs are $A_0^{\rm p}, \nu_1^{\rm p}, \nu_2^{\rm p}$ for proton,  and $A_0^{\rm e^-}, \nu_1^{\rm e^-}, \nu_2^{\rm e^-}$ for electron. We still allow for the fluctuation of positron background, which is denoted by a multiplier $c^{\rm e^+}$\cite{2015PhRvD..91f3508L}. For the local SNR, the proton and electron parameters are respectively $q_0^{\rm p}$, $\alpha^{\rm p}$, $E^{\rm p}_{\rm cut}$ and $q_0^{\rm e}$, $\alpha^{\rm e}$. Moreover, we have another parameter specifying the collision duration of CRs within the MC, $t_{\rm col}$. For the heavier nuclei, like C, N, O, they share the same power-law index of injection as proton. In addition, to fit the low energy data, we have to take solar modulation into account, which is represented by the modulation potential $\phi$\cite{2002astro.ph.12111M}. The parameters of transport, background sources and local SNR are summarized in the separated Tables \ref{tab:para_trans}, \ref{tab:para_bg} and \ref{tab:para_local}.
\subsection{Energy spectra}
\begin{figure*}
\centering
\includegraphics[height=20.cm, angle=0]{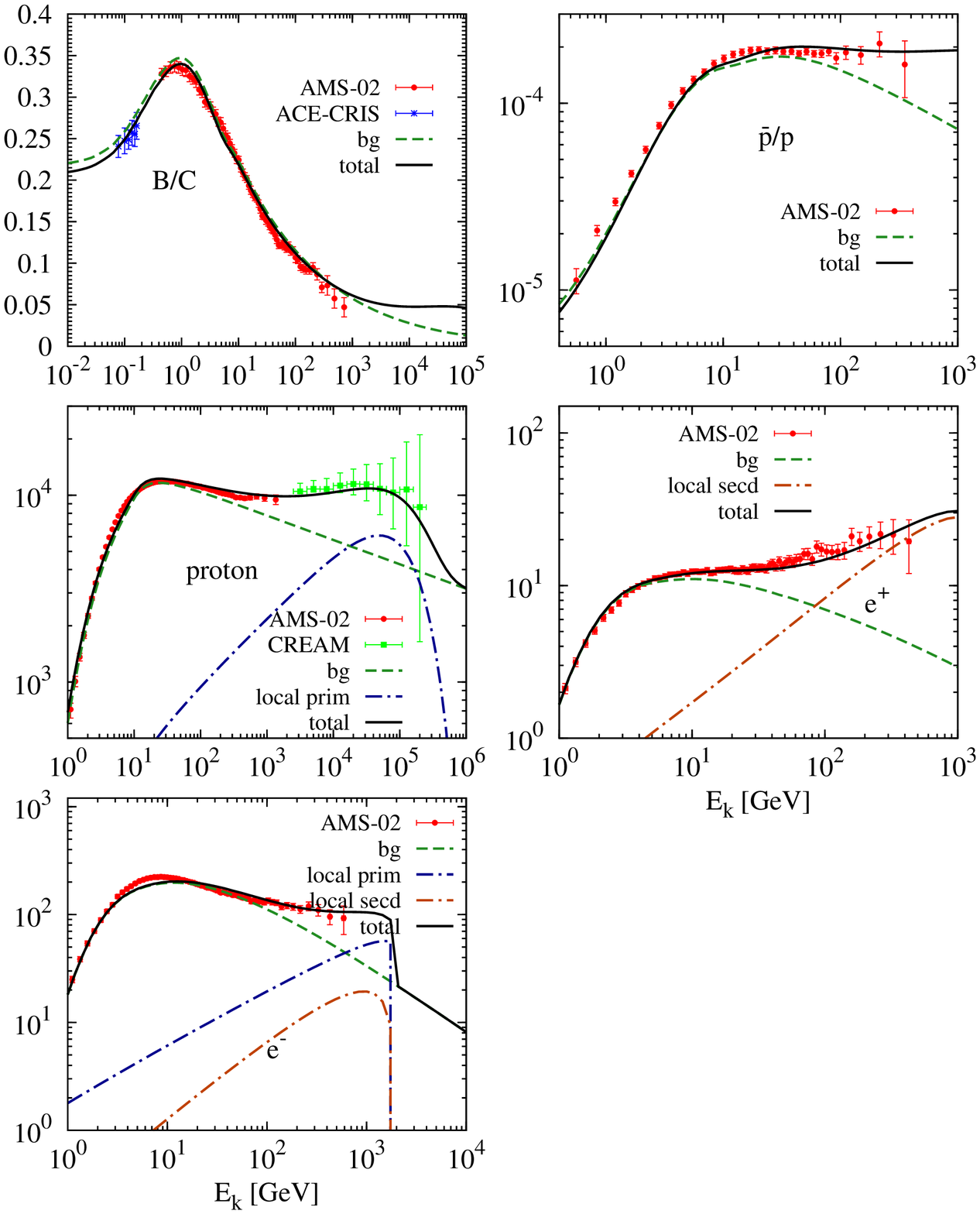}
\caption{
The fittings to CR energy spectra, $\rm B/C$(top left), $\rm \bar{p}/p$(top right), proton flux(middle left), positron flux(middle right) and electron flux(bottom left),  by DR-A model. For $\rm B/C$ ratio, both AMS-02(red) and ACE(1998/01-1999/01)(blue)\citep{2006AdSpR..38.1558D} data are used. The blue dash-dot is the flux from local SNR. The green dash and violet solid lines are respectively the background and total flux(or ratio) after solar modulation. The parameters of transport, background and local SNR are respectively reproduced in table \ref{tab:para_trans}, \ref{tab:para_bg} and \ref{tab:para_local}.  For $\rm B/C$, $\rm p$ and $\rm e^{\pm}$, the parameter of solar modulation $\phi_{\rm B/C} = 330$ MeV, $\phi_{\rm \bar{p}/p} = 660$ MeV and $\phi_{\rm e^{\pm}} = 1300$ MeV. 
}
\label{fig:spec_DRA}
\end{figure*}

Figure \ref{fig:spec_DRA} illustrates our fittings assuming a comparatively young local SNR, whose age is $10^5$ year old. In the figure, the fluxes/ratios from the background(green dash), the local primary protons and electrons(blue dash-dot), the local secondary electrons and positrons(brown dash-dot) and  total(black solid) are respectively reproduced. Both background and total fluxes/ratios have been solar-modulated to account for the low energy spectra. The modulation potential for $\rm p$($\rm \bar{p}$), $\rm e^{\pm}$ and $\rm B/C$ are respectively $\phi = 660, 1300$ and $330$ MeV. 

For B/C ratio, it can be seen that both background and total can well fit the AMS-02 data. In this energy range, the contribution from local SNR is negligible, and hence the differences between two of them are tiny. The required transport parameters are $D_{0} = 5.9 \times 10^{28}$ cm$^2$ s$^{-1}$, $\delta = 0.343$, $v_A = 30$ km s$^{-1}$ and $L = 4.85$ kpc, which are similar to other fittings\citep{2015PhRvD..91f3508L}. However above $600$ GeV, the local component starts to dominate which renders the total B/C ratio hardening. This character can be verified by the future observations. Something similar also happens in the $\rm \bar{p}/p$ ratio. From $\sim 10$ GeV, the total $\rm \bar{p}/p$ ratio gradually deviates from the prediction of conventional model and flattens up to $10^3$ GeV, which agrees well with AMS-02 data. 

Since both positrons and anti-protons are mainly generated by CR protons, the spectra of proton, $\rm \bar{p}/p$ and $\rm e^+$ collectively pin down the components of background and local proton. We find in order to fit these spectra all together, the power-index of local protons $\alpha = 1.837$ is harder than the background $\nu_2 = 2.49$. The high energy cutoff $E_{\rm cut}$ of local proton is adjusted to $1.4\times 10^5$ GeV for the sake of fitting high energy positron data. And to produce enough positrons at high energy, the duration of collision is $3.5\times 10^2$ years when the density of local MC $n = 1000$ cm$^{-3}$. 

For electron spectra, the local component comes from two origins, one is the shock-accelerated primary electrons within local SNR and the other is the products of the collision between CR nuclei and molecular gas. The normalization and injected power-law index of local primary electrons are respectively $q_0^{\rm e^-} = 10^{48}$ GeV$^{-1}$  and $\alpha = 2$. Therefore the electron to proton ratio in local SNR is smaller than $K_{\rm ep} = 0.01$.  And the injected spectrum is also harder than background electron $\nu_2 = 2.83$.

\begin{figure*}
\centering
\includegraphics[height=20.cm, angle=0]{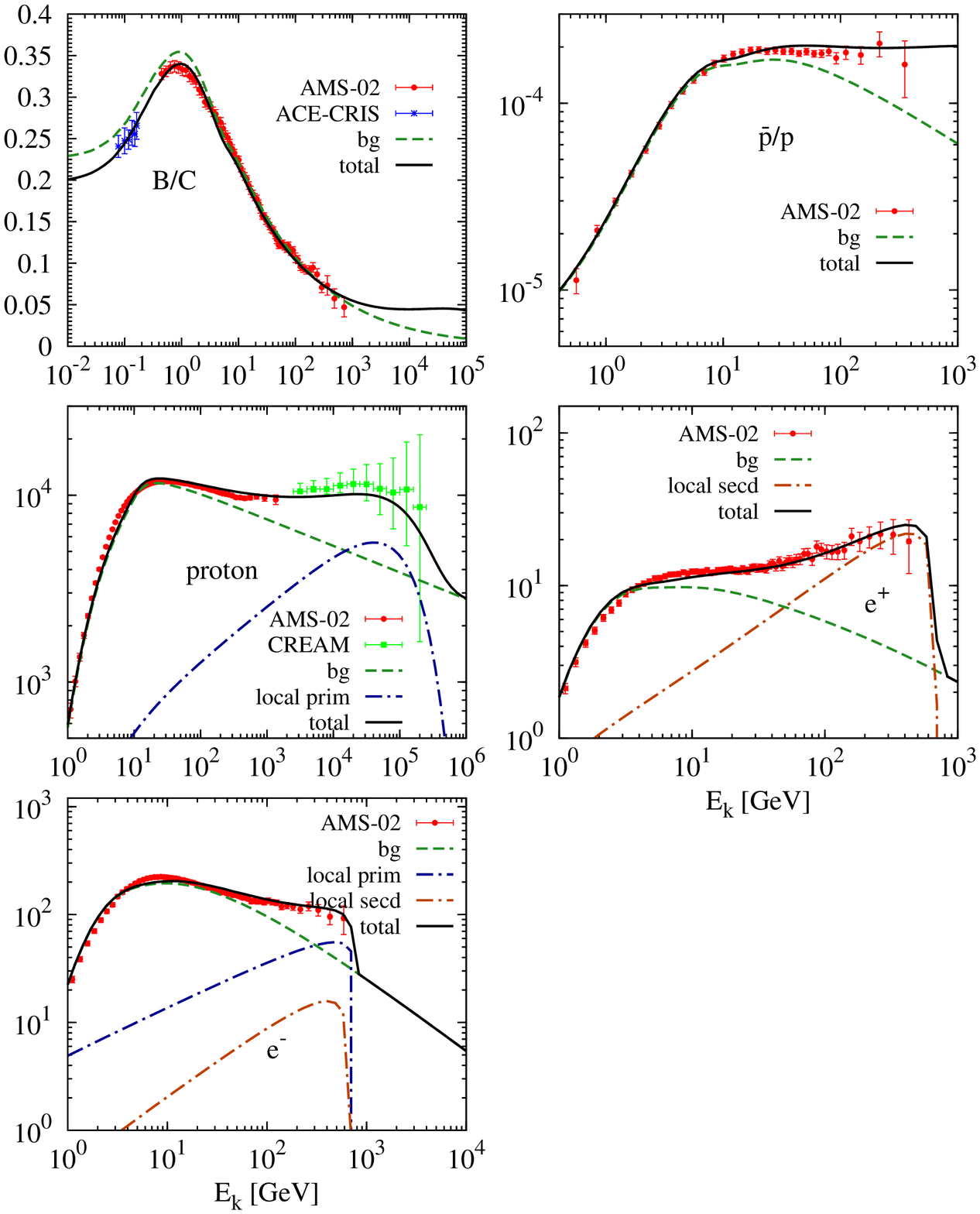}
\caption{
The fitting to CR energy spectra, B/C(top left), $\rm \bar{p}/p$(top right), proton flux(middle left), positron flux(middle right) and electron flux(bottom left),  under DR-B model. For B/C ratio, both AMS-02(red) and ACE(1998/01-1999/01)(blue)\citep{2006AdSpR..38.1558D} data are used. The blue dash-dot is the flux from local SNR. The green dash and black solid lines are respectively the background and total flux(or ratio) after solar modulation.  The parameters of transport, background and local SNR are respectively reproduced in table \ref{tab:para_trans}, \ref{tab:para_bg} and \ref{tab:para_local}. For $\rm B/C$, $\rm p$ and $\rm e^{\pm}$, the parameter of solar modulation $\phi_{\rm B/C} = 330$ MeV, $\phi_{\rm p} = 600$ MeV and $\phi_{\rm e^{\pm}} = 1300$ MeV.
}
\label{fig:spec_DRB}
\end{figure*}

Figure \ref{fig:spec_DRB} shows the similar results but hypothesizing an older local SNR(DR-B), whose age is fitted to be $3\times 10^5$ years old. Both fitted transport and background parameters are close to DR-A, but the injected power of local SNR has to be boosted to $q_0 = 95\times 10^{49}$ GeV$^{-1}$, nearly seven times higher than DR-A. The total energy spectra of $\rm B/C$ as well as $\rm \bar{p}/p$ resembles the model DR-A. Setting against Figure \ref{fig:spec_DRA}, the major differences come from the spectra of electron and positron, in which the total flux plunges dramatically at lower energy, around hundreds of GeV. Because of earlier releasing time of local CRs and energy loss during their propagation, only lower electrons are observed at present. This can be validated by the DAMPE experiment\cite{ChangJin:550}.

\subsection{Anisotropy}
As exemplified above, both local SNRs with different ages can well explain the current data. In order to distinguish the two cases, we further compute their accompanied anisotropies of protons and electrons. The dipole anisotropy is defined as
\begin{equation}
\delta = \frac{3 D_{xx}}{v} \frac{|\nabla \varphi|}{\varphi} .
\end{equation}
Owning to the spatial distribution of Galactic SNRs, there is inevitably a radial gradient of CR density, whose direction is toward Galactic edge. Therefore in the scenario of steady-state propagation, the anisotropy grows with diffusion coefficient, namely rises with the energy. However this is incompatible with current observations, e.g. proton\citep{2012JCAP...01..011B}, which do not manifest obvious energy-dependence between $1$ TeV and $100$ TeV. One of the interpretations is that there is a fresh nearby SNR(or more) lying at the anti-galactic center, whose CRs can offset the background streaming.

\begin{figure*}
\centering
\includegraphics[height=6.cm, angle=0]{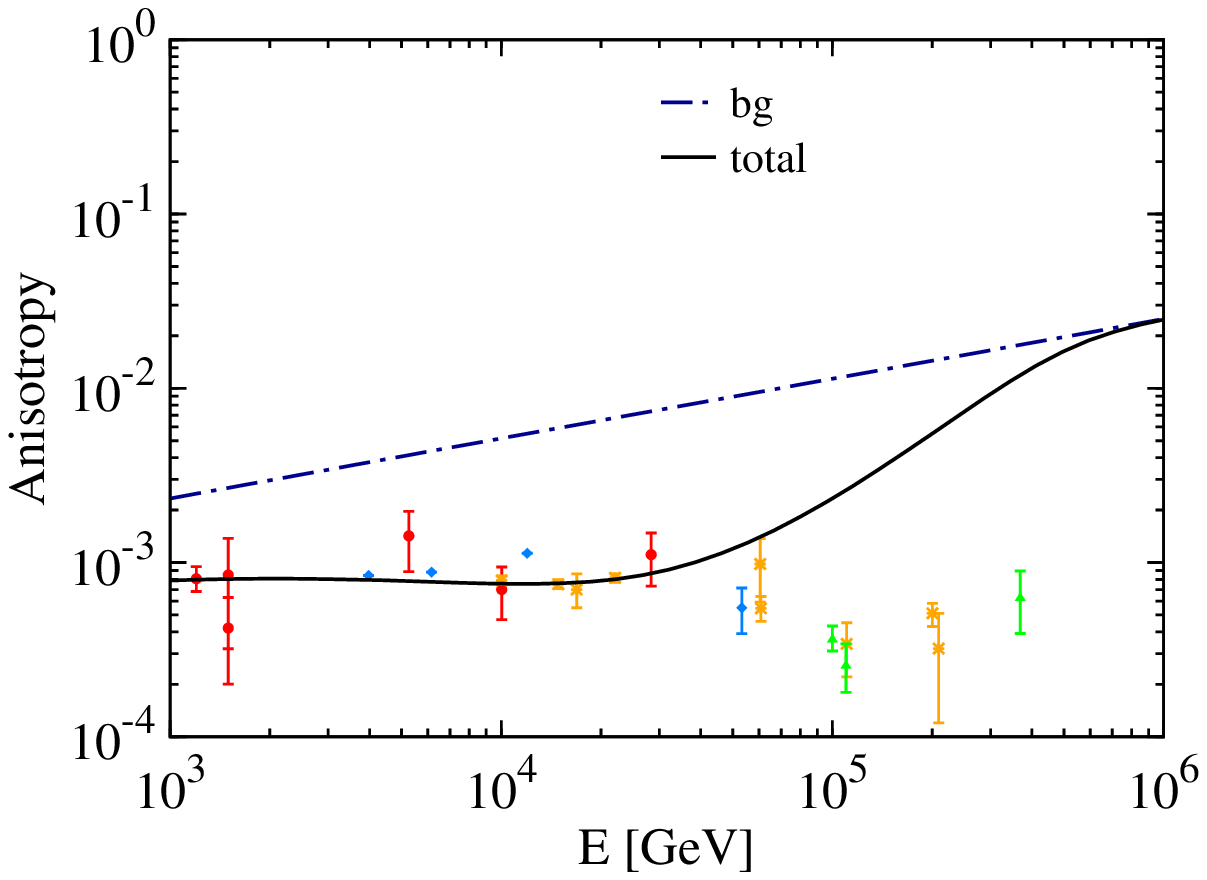}
\includegraphics[height=6.cm, angle=0]{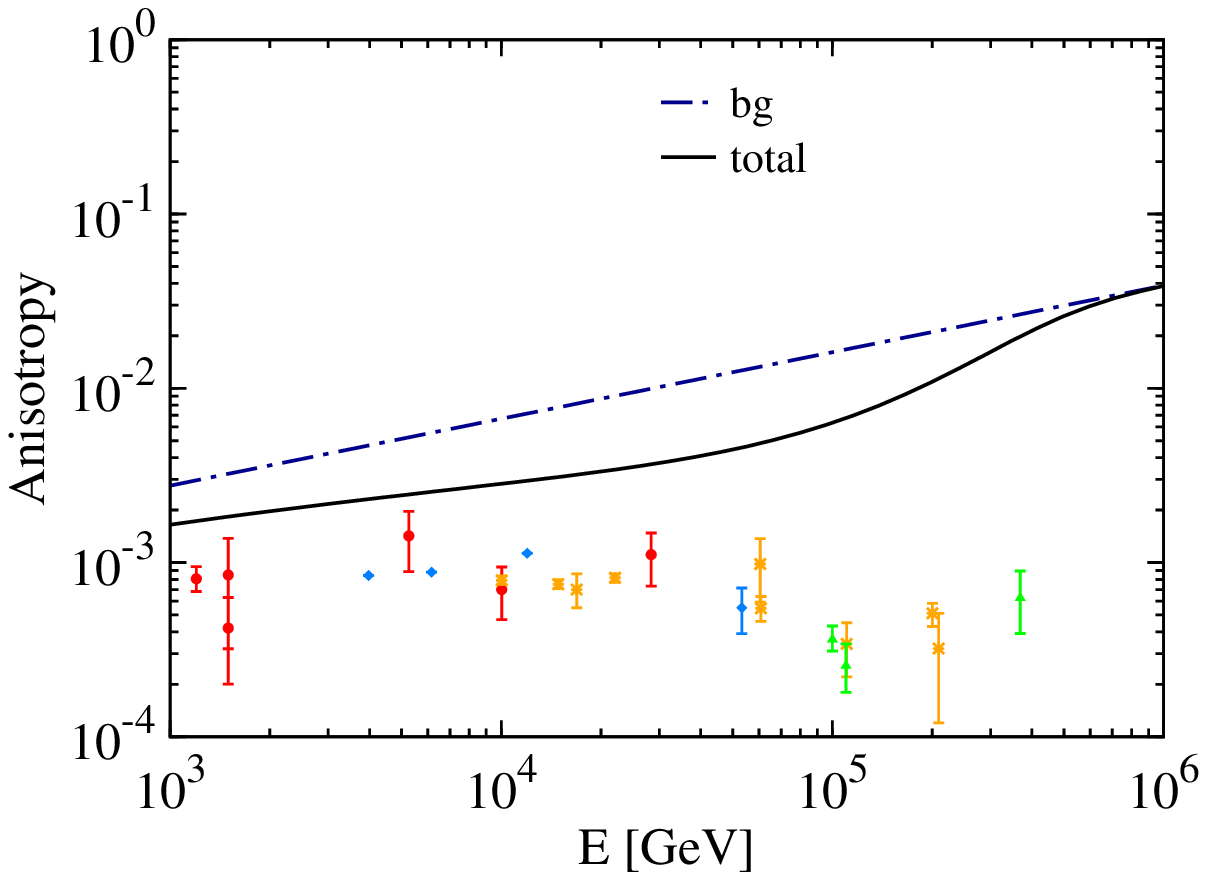}
\caption{
The anisotropy of $\rm p$ under DR-A(left) and DR-B(right) models. The blue dash-dot line is under the assumption of  continuous distribution of the sources and steady-state propagation, while the black solid line is the total one when considering the local SNR. The data points are taken from \citep{2005ApJ...626L..29A, 2009ApJ...692L.130A}. 
}
\label{fig:ani_prot}
\end{figure*}

In Figure \ref{fig:ani_prot}, we show the anisotropies of protons respectively in DR-A(left) and DR-B(right). The blue dash-dot line is evaluated from the background and the black solid line is the total one while adding the local SNR. Apparently, the nearby source at the anti-galactic center direction can effectively lower the anisotropy, but the extent of drop-off depends on the local SNR's age. For the younger one, the total anisotropy can well conform with the measurements from 1 TeV to tens of TeV. But above $100$ TeV, due to the high energy cut-off of local proton flux, the anisotropy comes back to the case of steady-state.

\begin{figure*}
\centering
\includegraphics[height=6.cm, angle=0]{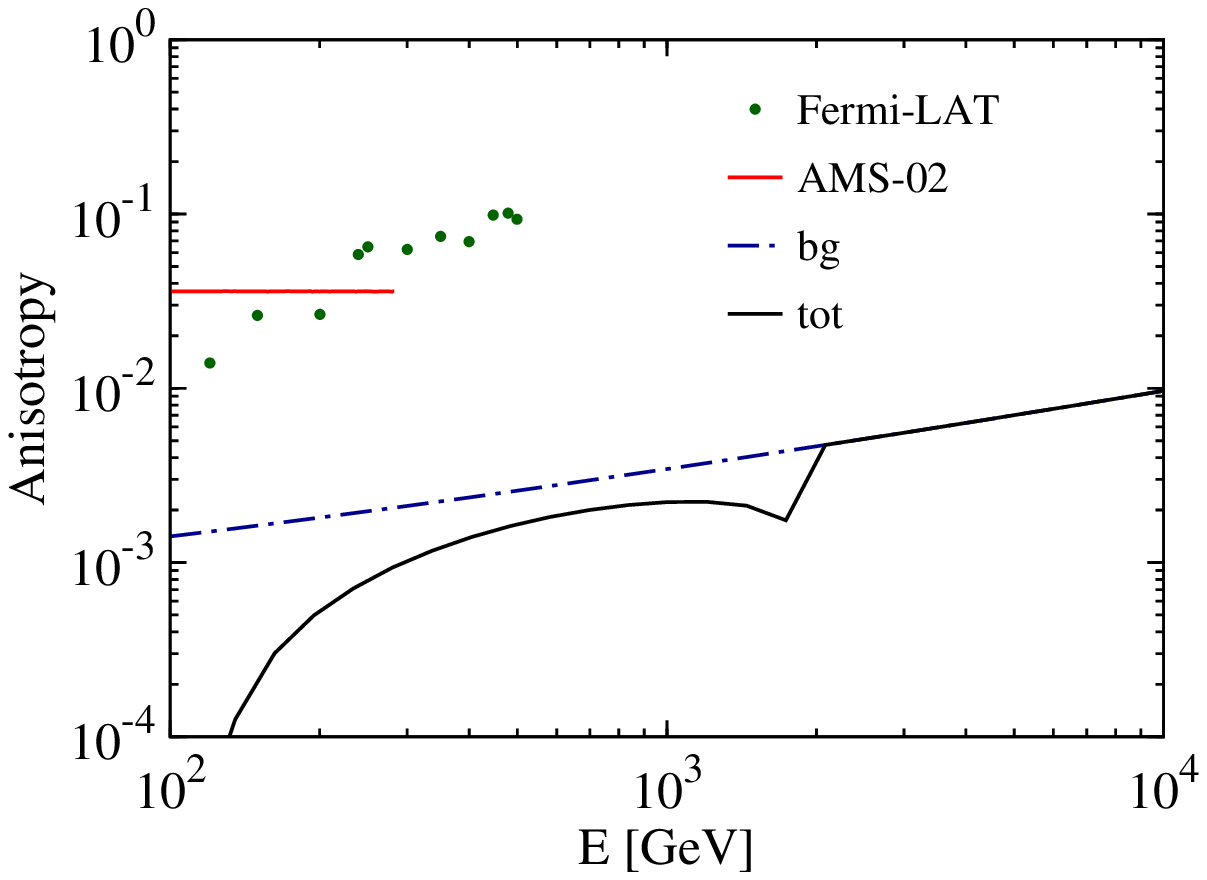}
\includegraphics[height=6.cm, angle=0]{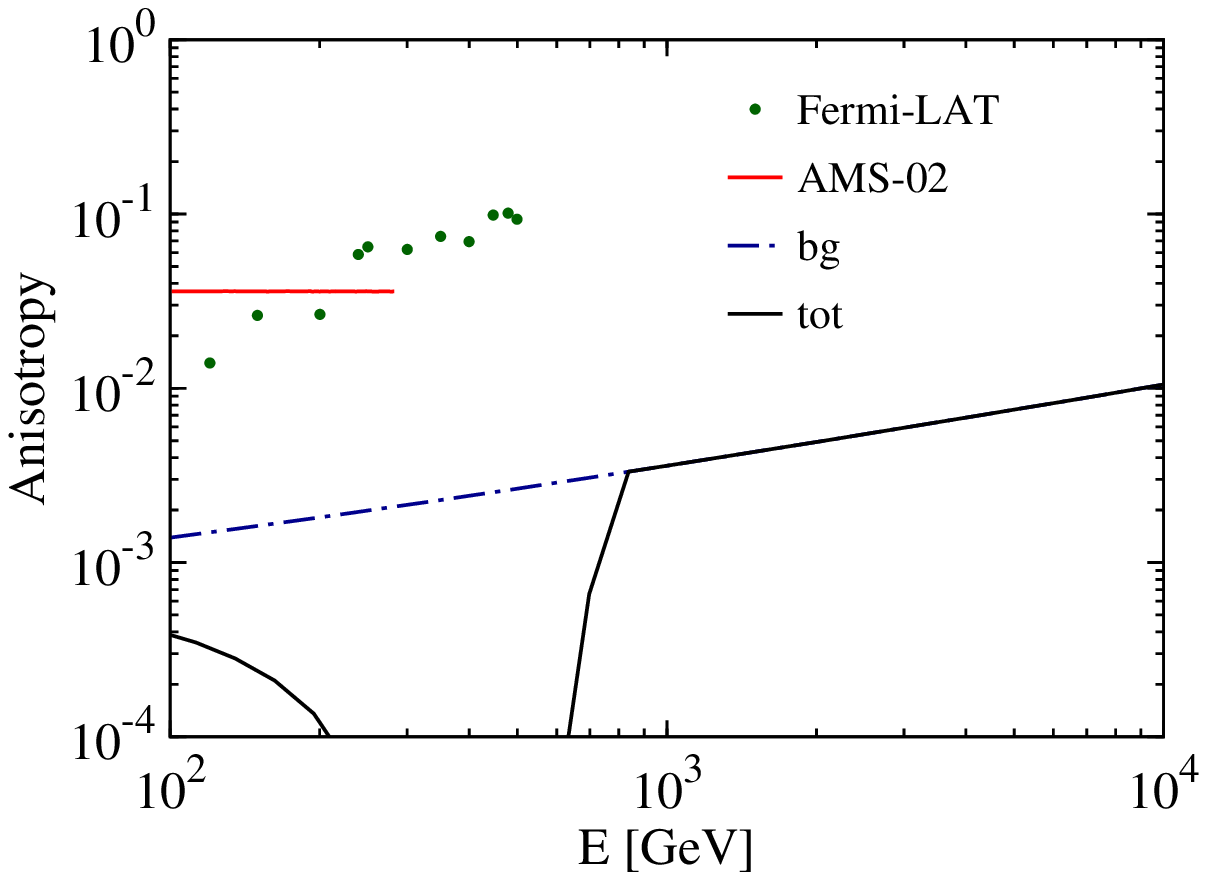}
\caption{
The anisotropy of $\rm e^{\pm}$ under DR-A(left) and DR-B(right) models. The blue dash-dot line is under the assumption of  continuous distribution of the sources and steady-state propagation, while the black solid line is the total one when considering the local SNR. The black dots are the limits from $1$ year  Fermi-LAT data\citep{2010PhRvD..82i2003A} and the red line is the recent AMS-02 limits on $\rm e^{\pm}$ anisotropy\citep{2013PhRvL.110n1102A}.
}
\label{fig:ani_lept}
\end{figure*}

Due to the relatively close origin, the discreteness of the sources is more important for the anisotropy of electrons. However for the electrons, current measurements only give the upper bounds\citep{2010PhRvD..82i2003A, 2013PhRvL.110n1102A}. We compute the influence of anisotropies under both local source models, as shown in Figure \ref{fig:ani_lept}. The blue dash-dot and black solid lines are respectively the anisotropies from the background assuming  continuous distribution of the background sources and the total. Compared to DR-A, the local streaming in DR-B is comparable to the background and the anisotropy at hundreds of GeV is tremendously suppressed. Nevertheless both anisotropies are still much less than the current constraints from  $1$-year Fermi-LAT\citep{2010PhRvD..82i2003A} and AMS-02\citep{2013PhRvL.110n1102A}.
%
\begin{table*}[t!]
\begin{center}
\begin{tabular}{ccccccc}
\toprule[1.5pt]
model  & $D_0$ [$10^{28}$cm$^2$/s] & $\delta$ & ${\cal R}_0$ [GV] & $v_A$ [km/s] & $L$ [kpc] \\
\hline
DR-A   & 5.9  & 0.343  & 4 & 30 & 4.85 \\
DR-B   & 5.5  & 0.383  & 4 & 30 & 4.85  \\
\bottomrule[1.5pt]
\end{tabular}
\end{center}
\caption{Set of transport parameters.}
\label{tab:para_trans}
\end{table*}

%
\begin{table*}[t!]
\begin{center}
\begin{tabular}{cccccccccccc}
\toprule[1.5pt]
model  & $\nu_1^{\rm p}$ & ${\cal R}_{\rm br}^{\rm p}$ [GV] & $\nu_2^{\rm p}$ & $E_{k0}^{\rm p}$ [GeV] & $A_0^{\rm p}$ & $\nu_1^{\rm e^{-}}$ & ${\cal R}_{\rm br}^{\rm e^{-}}$ [GV] & $\nu_2^{\rm e^{-}}$ & $E_{k0}^{\rm e^{-}}$ [GeV] & $A_0^{\rm e^{-}}$ \\
\hline
DR-A   & 1.81 & 12.88 & 2.49  & 100 & 4.16  & 1.79  & 2.74  & 2.83  & 25 & 15.4  \\
DR-B   & 1.70 & 12.88 & 2.46  & 100 & 4.05  & 1.70  & 2.74  & 2.88 & 25  & 14.45 \\
\bottomrule[1.5pt]
\end{tabular}
\end{center}
\caption{Sets of parameters of background CR injection and solar modulation. The normalization flux of proton $A_0^{\rm p}$ is in units of $10^{-9}/({\rm cm}^2\cdot {\rm sr}\cdot {\rm s}\cdot {\rm MeV})$ and the electron's $A_0^{\rm e^{-}}$ is $10^{-10}/({\rm cm}^{2}\cdot {\rm sr}\cdot {\rm s}\cdot {\rm MeV})$.}
\label{tab:para_bg}
\end{table*}


%
\begin{table*}[t!]
\begin{center}
\begin{tabular}{cccccccccc}
\toprule[1.5pt]
model & $\alpha^{\rm p}$ & $q_0^p$ [$10^{49}$ GeV$^{-1}$] & ${\cal R}_{0}$ [GV] & $E^{\rm p}_{\rm cut}$ [GeV] & $\alpha^{\rm e}$ & $q_0^e$ [$10^{49}$ GeV$^{-1}$] & $d$ [kpc] & $t_{\rm age}$ [$10^5$ yr] & $t_{\rm col}$ [$10^2$ yr] \\
\hline
DR-A   & 1.837  & 14    & 12.88 & 1.4e5 & 2  & 0.1 & 0.1 & 1  & 3.5 \\
DR-B   & 1.837  & 100  & 12.88 & 1.4e5 & 2  & 1 & 0.1 & 3  & 3.5 \\
\bottomrule[1.5pt]
\end{tabular}
\end{center}
\caption{the parameters of local young SNR. $\alpha^{\rm p}$ and $\alpha^{\rm e}$ are the injected power-index for proton and electron. $q_0^{\rm p}$ and $q_0^{\rm e}$ are respectively the source term of proton and electron at rigidity ${\cal R}_0 = 12.88$ GV. $d$ is its distance to solar system and $t$ the injection time of CR. $t_{\rm col}$ is the interaction time between CRs and molecular gas.}
\label{tab:para_local}

\end{table*}
%

\section{Conclusion} \label{sec:concl}

Currently, the excesses seem prevalent in both primary and secondary  galactic CRs, according to the recent observations. In this work, we address ourselves to this universal anomaly in a unified viewpoint. Specifically, we envisage there was once a nearby supernova explosion, which occurred within the giant molecular cloud. The primary CRs are accelerated by the shock wave. And the secondaries, $\rm e^{\pm}$, $\rm \bar{p}$ and $\rm B$ are massively produced by the interactions between primary CRs and molecular gas. All of them get free and enter into the interstellar space once the molecular cloud is fragmented. 

In order to fit the spectra of proton, $\rm \bar{p}/p$ ratio and positron, the power index of local source has to be harder than the average of background SNRs.  For the $\rm B/C$ ratio, the low energy spectrum is unaffected by the local source. However beyond $1$ TeV, the total B/C ratio is well above the background. This is different from the spectrum of $\rm \bar{p}/p$ ratio, whose transition occurs at lower energy. This stems from the differences of production cross section between boron and antiproton. Future measurements extending to higher energies can verify this notable feature. 

For the local SNR, we hypothesize two different ages, both of which can well recreate all the spectra. The main difference is at the high energy cut-off in the spectra of electron and positron. For the younger SNR, the cut-off occurs above $1$ TeV, while for the older, the spectrum falls off well before $1$ TeV, about hundreds of GeV. The ongoing DAMPE experiment\cite{ChangJin:550} could supply more precise measurement at this energy range.

For the local source model, one of the validation methods is to compare its anisotropy of high energy CRs with observations. Under steady-state assumption, the anisotropy is proportional to the diffusion coefficient, which is in conflict with the accumulative data. If a local SNR is located at anti-Galactic center direction, the CR flux from it could effectively counteract the streaming from the Galactic center. We compute the anisotropies under our local SNR models. We find that when the local SNR is younger, the total anisotropy of proton is in a good agreement with the observations from $1$ TeV to tens of TeV. In the case of electron, the local streaming cancels the background streaming in DR-B model and the anisotropy is greatly suppressed at hundreds of GeV. The anisotropies from both models are far below the current upper limits from AMS-02 and Fermi-LAT.

\begin{acknowledgments}
We would like to thank Qiang Yuan for helpful discussion and reading the manuscript.
This work is supported by the NSF of China under Grants Nos. 11475189, 11475191, and by the 973 Program of China under
Grant No.~2013CB837000, and
by the National Key Program for Research and Development (No. 2016YFA0400200).
\end{acknowledgments}

\bibliographystyle{unsrt_update}
\bibliography{ref}

\end{document}